\documentclass[conference]{IEEEtran}
\IEEEoverridecommandlockouts

\usepackage{textcomp}
\usepackage{xcolor}
\def\BibTeX{{\rm B\kern-.05em{\sc i\kern-.025em b}\kern-.08em
    T\kern-.1667em\lower.7ex\hbox{E}\kern-.125emX}}

\usepackage{blindtext}
\usepackage{graphicx}
\usepackage{wrapfig}
\usepackage{lipsum}
\usepackage{amssymb}

\ifCLASSINFOpdf
  \graphicspath{{images/}}
  \DeclareGraphicsExtensions{.pdf,.jpeg,.png}
\else
\fi

 \renewcommand{\baselinestretch}{0.93}
 \usepackage[font={small},labelfont=bf ]{caption}
 \usepackage[font=footnotesize]{subfig}
 \usepackage{amsmath,amsfonts,amsthm,bm}
 \usepackage{stfloats}
 \usepackage[font={small,it}]{caption}
 \usepackage{cite}
 \usepackage{tabularx}

\usepackage{pifont}
\usepackage{booktabs} 
\usepackage{multirow}
\usepackage{mathtools}
\usepackage{soul}
\usepackage{subfig}
\usepackage{algorithm}
\usepackage[]{algpseudocode}
\usepackage{algorithmicx}
\algnewcommand\algorithmicforeach{\textbf{for each}}
\algdef{S}[FOR]{ForEach}[1]{\algorithmicforeach\ #1\ \algorithmicdo}
\algtext*{EndWhile}
\algtext*{EndIf}
\algtext*{EndFor}
\usepackage{graphicx}
\usepackage{xcolor}

\begin{document}

\title{ \vspace{-2mm} LASCA: Learning Assisted Side Channel Delay Analysis for Hardware Trojan Detection \vspace{-3mm}\\
\thanks{This research is funded by the Defense Advanced Research Projects Agency (DARPA-AFRL, FA8650-18-1-7820) of the USA.}
}

\author{
    \IEEEauthorblockN{
        Ashkan~Vakil\IEEEauthorrefmark{1}, 
        Farnaz~Behnia\IEEEauthorrefmark{1}, 
        Ali~Mirzaeian\IEEEauthorrefmark{1}, 
        Houman~Homayoun\IEEEauthorrefmark{2},
        Naghmeh~Karimi\IEEEauthorrefmark{3}, 
        Avesta~Sasan\IEEEauthorrefmark{1}}
    \IEEEauthorblockA{
        \IEEEauthorrefmark{1}Department of ECE, George Mason University, e-mail: \{avakil, fbehnia, amirzaei, asasan\}@gmu.edu \\
        \IEEEauthorrefmark{3}Department of CSEE, University of Maryland, Baltimore County, e-mail: nkarimi@umbc.edu \\            
        \IEEEauthorrefmark{2}Department of ECE, University of California, Davis, e-mail: hhomayoun@ucdavis.edu \vspace{-4mm}
    }
}

\maketitle
\thispagestyle{plain}
\pagestyle{plain}

\begin{abstract}
In this paper, we introduce a Learning Assisted Side Channel delay Analysis (LASCA) methodology for Hardware Trojan detection. Our proposed solution, unlike the prior art, does not require a Golden IC. Instead, it trains a Neural Network to act as a process tracking watchdog for correlating the static timing data (produced at design time) to the delay information obtained from clock frequency sweeping (at test time) for the purpose of Trojan detection. Using the LASCA flow, we detect close to 90\% of Hardware Trojans in the simulated scenarios.
\end{abstract}

\section{Introduction and Background} \label{intro}
The use of untrusted entities in this global supply chain has raised pressing concerns about the security of the fabricated ICs that are targeted for use in sensitive applications. One of these security threats is the adversarial infestation of fabricated ICs with a hardware (HW) Trojan. A Trojan can be broadly defined as a malicious modification to a circuit to control, modify, disable, or monitor its logic.

Conventional manufacturing VLSI test and verification methodologies fall short in detecting HW Trojans due to the different and un-modeled nature of these malicious alterations. This has led many researchers to investigate solutions for detection of HW Trojans through statistical analysis of side-channel information collected from ICs, including side-channel power analysis \cite{agrawal2007trojan, rad2010sensitivity, salmani2009new, lecomte2017chip, liu2014hardware, lamech2011rebel}, power supply transient signal analysis \cite{rad2008sensitivity, ra2008power}, regional supply currents analysis \cite{du2010self}, temperature analysis \cite{hu2013high}, wireless transmission power analysis \cite{liu2013hardware}, and side-channel delay analysis \cite{ 6472276,4559049,li2008speed, jin2008hardware, cui2018hardware, exurville2015resilient, ismari2016detecting}.

The problem with many of the previous HW Trojan detection solutions is a need for some sort of a golden model from which the parametric signature of the fabricated ICs are collected and used to define a decision boundary (power, delay, temperature, etc) for separating the Trojan-infested ICs. However, building a golden IC is extremely difficult or even impossible: In many cases, especially in advanced technology node, the choice of the foundry is limited to one or a very few, none of which may be trusted. Even if a trusted foundry exists, fabrication of a small volume of ICs for obtaining a golden IC is usually cost prohibitive\cite{liu2014hardware}. Moreover, the process used in each foundry is quite different and a golden IC that is fabricated in one foundry can not be used for assessing an IC fabricated in another foundry.

For these reasons, we do not assume the existence of a golden IC or a golden model. Instead, we develop and train a learning-assisted timing-adjustment model that combined with the STA acts as a golden model. This work is motivated by two previous papers: The side-channel power analysis in \cite{liu2014hardware} and side-channel delay analysis in \cite{6472276}, a short description of which is given next:

The side-channel statistical power analysis solution for Trojan detection in \cite{liu2014hardware} proposed that the trusted region for the operation of a Trojan free IC can be learned using a combination of a trusted simulation model, measurements from the carefully engineered and distributed PCM structures, and advanced statistical tail modeling techniques. This work, however, relies on side-channel power analysis for the detection of hardware Trojan. For observing a meaningful change in leakage or dynamic power, the size of hardware trojan has to be large. Hence, this technique falls short of detecting Hardware Trojans implemented using a small number of gates. This is when our proposed solution can detect even a single added logic gate in a tested timing path. Besides, \cite{liu2014hardware} relies on the usage of PCMs (with a defined structure that is repeated and distributed over the IC) for extracting the process parameters. However, the number and accuracy of PCMs are limited. Although PCM can roughly track the process corner from chip to chip and could be used for the rough calibration of timing and spice models, they fall short of accurately characterizing the behavior of different gates and metal layers. This is when in our proposed solution, every timing path could be used as a PCM for training the neural assisted timing augmentation engine, and therefore the impact of different timing path topologies, different gate types/sizes, and the change in the capacitive or resistive load of different metal layers are taken into account.  

The side channel delay analysis solution in \cite{6472276} uses Clock Frequency Sweeping Test (CFST) to detect the hardware Trojan. However, it relies on the existence of a Golden IC for delay comparison. Our proposed side-channel Trojan detection scheme is inspired by this work (and used CFST for the generation of label data points for each feature set), however, our proposed mechanism does not need a Golden IC. 

\vspace{-1mm}
\section{Trojan Threat Model} 
\begin{figure}[t]
    \centering
    \includegraphics[width=\columnwidth]{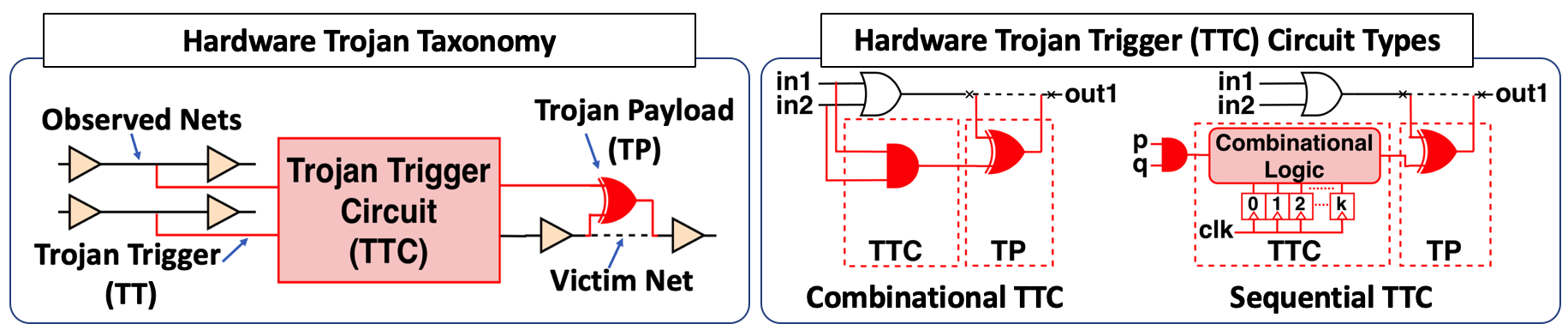}
    \caption{(left): Trojan taxonomy, (right): Trojan trigger circuit types}
    \label{trojan_terminology}
\end{figure}

\vspace{-1mm}
The adversary in this paper is an untrusted foundry with access to GDSII (Graphic Database System format). The goal of the adversary is to insert a Trojan that is triggered based on a combination, or a sequence of rare events. A Trojan, As illustrated in Fig. \ref{trojan_terminology}, consists of 1) Trojan's Trigger inputs (TT), 2) Trojan's Triggering ( which could be sequential or combinational) Circuit (TTC), and 3) Trojan Payload (TP). Upon activation, the TP alters the circuit functionality. We assume that no Golden IC exists, and the Trojan is inserted in all fabricated ICs.

\section{Trojan Detection Challenge: Variability} \vspace{1mm}
The TT of an HW Trojan poses an additional capacitive load on its driving cell, resulting in a slower rise and fall, while its TP adds a gate delay to its victim timing path. In a perfect world, A Trojan can be detected by tracking and analyzing the changes in the delay of timing-paths compared to that predicted by STA. The challenge for this solution is that STA suggested delay information can be significantly different from delay information that is collected at the test time.  This is due to several factors most notable of which are:  1) voltage noise, 2) Process Variation,  and 3) process drift. \\[-8pt]

\textbf{1- Voltage Noise:} In an ASIC chip, the Power Delivery Network (PDN) is an RLC network that responds to the change in the current demand of transistors, posing voltage (IR) drop and voltage variation on transistors \cite{ASPDAC}. During STA the IR drop and voltage noise are modeled by (1) specifying a rail voltage value below supplied voltage to account for IR drop, and (2) using register-endpoint uncertainty to guard against voltage-variation-induced clock network jitter \cite{arabi2007power}. The chosen values for the rail-voltage and uncertainty should be pessimistic to capture the worst-case (to prevent setup/hold failure). However, the majority of timing-paths experience smaller IR-drop and voltage noise. This poses a security threat; the pessimistic margins build large unused timing slack into the majority of timing-paths, which is not visible to the physical designer and test engineer. The unused timing slacks can be used by an adversary in an untrusted foundry to design a Trojan and hide its delay impact. \\[-8pt]

\textbf{2- Process Variation (PV):} The PV refers to the variations in the physical and electrical properties of transistors in the result of physical limitations faced during the fabrication process \cite{4479829}. It affects the delay and drive-strength of fabricated transistors. PV makes Trojan detection more difficult as one has to investigate if the change in the delay is the result of PV or the timing impact of an HW Trojan. \\[-8pt]

\textbf{3- Process Drift:} The SPICE model for the fabrication process in a new technology node is released soon after a process is production-ready and is used to characterize the standard cell libraries deployed in a physical design house. To guarantee a high-yield, the SPICE-model and standard cell libraries are padded with a carefully crafted margin. In addition, the foundry keeps improving the process over time to increase yield and reduce cost. Hence, the fabrication process and the released SPICE model \textit{drift} apart over time. The improvement in the process builds large unused slacks in a fabricated IC that is designed using the older SPICE and library models. This poses a security problem as these unused and hidden timing slacks can be used by an adversary in the untrusted foundry to design stealthy HW Trojan(s).

\section{Proposed HW Trojan Detection Solutions}
The LASCA integrates multiple variation modeling and mitigation techniques into a side-channel delay analysis solution for the purpose of HW Trojan testing. Using the proposed techniques, we characterize and mitigate the impact of voltage noise, PV and process drift to improve the correlation between the adjusted timing model and the fabricated ICs' timing behavior. We first describe how each of these variation sources is modeled and mitigated, and then explain how each mitigation technique is integrated into LASCA to improve the chances of an HW Trojan detection.

\subsection{Modeling and Tracking the Process Drift}\label{NN_section}

Process drift results in a non-uniform shift in the delay of different timing-paths. To model the timing impact of process drift, we design and train a Neural Network (NN)to act as a process tracking watchdog (NN-Watchdog). This NN-Watchdog is used to predict the difference between the slack reported by STA at design time and that sampled from fabricated IC at test time. To train the NN-Watchdog, we need a labeled data-set. Each data point in our data-set is a collection of 48 input features and a label value. The input features, detail of which is in Table \ref{feature_table}, are extracted from physical design EDA and the STA engine.  The label for each data point is the difference between the slack reported by STA (at design time), and that obtained by CFST \cite{6472276} (at test time).

\begin{table}[t]
 \vspace{2mm}
\centering
\caption{Description for each of 48 features, extracted from each timing-path for building the NN training set. (LP: Launch portion of timing-path, CP: Capture portion of timing-path, DP: Data portion of timing-path, M: Metal Layer, x: drive strength of the gate)}
\label{feature_table}
\scriptsize
\centering
\scalebox{0.9}{
\begin{tabular}{| l | l |l | l | }
\hline
\multicolumn{3}{|l|}{\textbf{Total of 48 Features, 3 Feature Extracted from each timing-path}} \\
\hline 
  Setup Time 	&	 Path delay reported in STA       	&	 Sum of fanout over cells in DP  \\
 \hline
\multicolumn{3}{|l|}{\textbf{45 Feature Extracted, 15 from each sub-path (CP, LP and DP) }} \\
\hline 
 number of gates 	&	subpath Delay    	&	 \# cells of x0 strength  \\
 \hline
  \# cells of x1 strength 	&	\# cells of x2 strength   	&	 \# cells of x4 strength  \\
 \hline
   \# cells of x8 strength 	&	\# cells of x16 strength   	&	 \# cells of x32 strength  \\
 \hline
	Total Length of M1 		&	Total Length of M2   	&	    Total Length of M3     	    	\\
  \hline
   	Total Length of M4 	    &	Total Length of M5  	&	 Total Length of M6   		    	\\
  \hline
\end{tabular}						
}
\normalsize
\end{table}

To assess the effectiveness of NN-Watchdog (and for lack of access to fabricated ICs), we modeled the process drift by extracting the shift in delay values from SPICE simulations performed using a skewed SPICE model. For this purpose, we first extracted the SPICE model for each timing-path in the input training. Then, to mimic a systematic process drift, the SPICE model was skewed such that the NMOS and PMOS transistors were $\sim$X\% faster, and the Metal capacitance for Metal layers 1 to 7 was derated by Y\%. Selection of $X$ and $Y$ gives us a consistently faster or slower process model. For example, the selection of $(X,Y) = (5,5), (0,0), (-5,-5)$ produces Fast, Typical, and Slow process models in our simulations. The resulting database was then divided into 1) training-set for training the NN (60\% of timing-paths), 2) verification-set used for assessing the trained model accuracy while training (20\% of timing-paths), and 3) test-set used for reporting the results (20\% of timing-paths).

\begin{figure}[t]
\centering
  \includegraphics[width=\columnwidth]{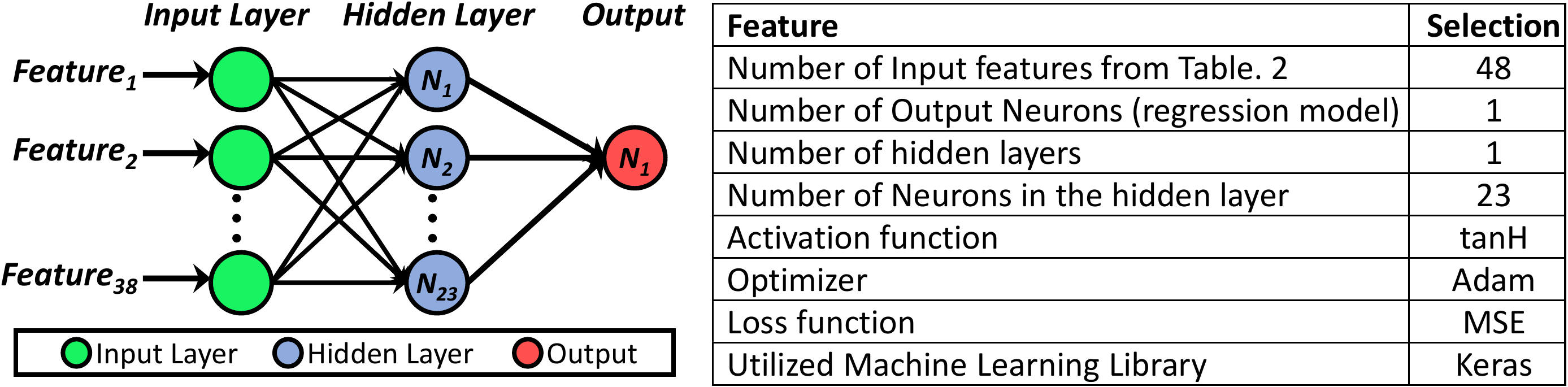}
  \caption{The configuration of the feed-forward fully-connected NN trained in this work to serve as a test-time process watchdog.}
 \label{regmodel}
 \vspace{1mm}
\end{figure}

We then design and trained a fully connected feed-forward NN (Fig.~\ref{regmodel}.(left)) as a process tracking watchdog that predicts the difference between the slack reported by STA and slack measured by the tester. To find a NN architecture with high accuracy, we utilized Keras \cite{chollet2015keras} and trained a large number of models by sweeping various model parameters. The number of hidden layers was swept between 1 to 3, and the number of nodes in each hidden layer was swept from $[{\frac{input+output}{2}}$ to ${\frac{2\times(input+output)}{3}}]$. We also tested different activation functions including: $tanH$, $Sigmoid$, $ReLU$, $PReLU$, $Power$, $Log$, and $Exp$. The object of the training was defined to reduce the sum of squared distances (MSE) between the model's (slack shift) prediction and labeled data.  

We separately trained each model for AES128, Ethernet, and S38417 (from IWLS benchmark suit\cite{iwls2}) benchmarks using 21K, 20K, and 4K timing-paths for training, respectively. Data collected from the training of over 5K models revealed that the configuration that is shown in Fig.~\ref{regmodel}.(right) achieves the highest regression accuracy in most cases. When using a single "NVIDIA Tesla k80" GPU, the training time of this network for s38417, Ethernet and AES128 was approximately 1,4 and 5 hours respectively.

\subsection{Modeling and Mitigating PV}

\begin{figure}[t]
\centering
\includegraphics[width=\columnwidth]{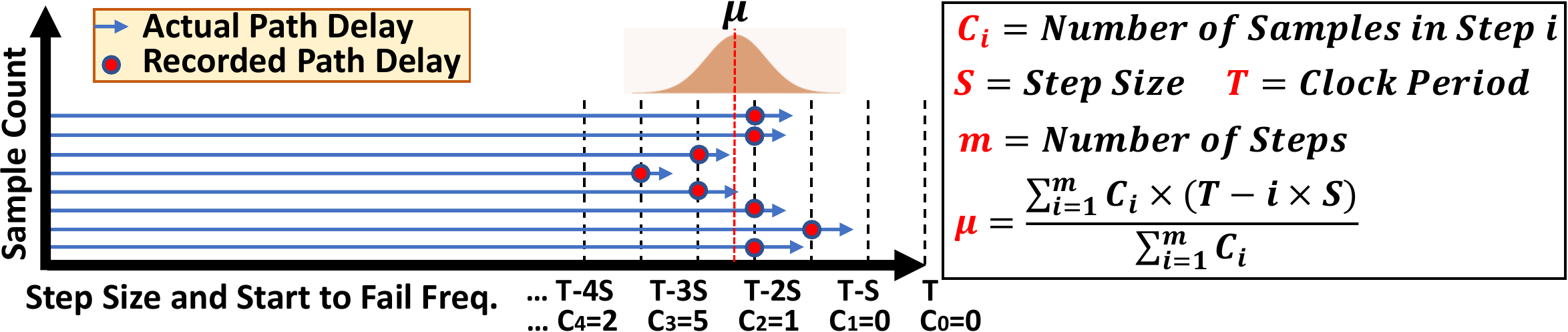}
\caption{Computing the mean delay of a path using CFST delay measurements with step size S, clock period T, over m samples (dies).}
\label{cancelling_process_variation}
\vspace{2mm}
\end{figure}

We divide the PV into two categories: 1) \textit{Random} Class that includes the independent intra-die PV, and 2) \textit{Persistent} class including all forms of inter-die and correlated intra-die variation.  
we use two different mechanisms to deal with random and persistent PV: (1) We perform speed binning on fabricated ICs and divided them into different speed bins (Fast, Normal, and Slow), arguing that ICs in the same bin are similarly affected by the persistent PV. Then for each bin, we train an NN-Watchdog. (2) To reduce the impact of random PV, using the formulation presented in Fig. \ref{cancelling_process_variation}, we collect the delay of each timing path (in our test set) from many ICs and compute their average delay to be used in our HW Trojan detection solution. When the timing-path delay is averaged across $N$ different dies, the standard deviation of the random variable representing the average delay is N times smaller than the standard deviation of individual samples ($\sigma_{AVG}  = \sigma_{sample}/N$). Note that the mean value is computed from discrete delay samples obtained from CFST test, and the tester's size (S), as illustrated in Fig. \ref{cancelling_process_variation}, affects the value of the computed mean.

To emulate the persistent PV (within the same process corner), we created 2 additional derivatives (slightly modified copy) for each of our skewed SPICE models. Each skewed SPICE model was altered to make the transistors in the first derivative 1\% slower, and in the second derivative 1\% faster. To model Random PV, each SPICE simulation is subjected to 100 Monte Carlo simulations (modeling CFST performed on 100 different dies in the same speed-bin), where the threshold voltage ($V_{th}$), Oxide thickness ($T_{ox}$) and channel Length ($L$) are varied (based on a normal distribution) to model the variation of path delays from chip to chip according to the expected variation in 32nm technology node.

\subsection{Modeling Timing Impact of Voltage Noise}
To improve the accuracy of our timing model (GTM), we utilize the IR-ATA methodology in \cite{ASPDAC} that models the voltage drop and voltage noise. The IR-ATA flow models the voltage drop and endpoint uncertainty (due to the voltage-induced clock jitter) using a differential voltage pair (different voltages for launch and capture path of a timing-path). The differential voltage pair is obtained based on a statistical analysis performed on the design-specific IR simulation results. By using the IR-ATA, the voltage-induced clock jitter uncertainty becomes path specific. This removes the need for a large hard margin, resulting in the majority of timing-paths to benefit from the smaller and dynamically computed margins. Due to lack of space, we avoid discussing the details of IR-ATA and refer the readers to \cite{ASPDAC} for more details.

\subsection{LASCA Trojan Detection Flow} \label{detection_flow_variation}
\begin{figure}[t]
\centering
\includegraphics[width=\columnwidth]{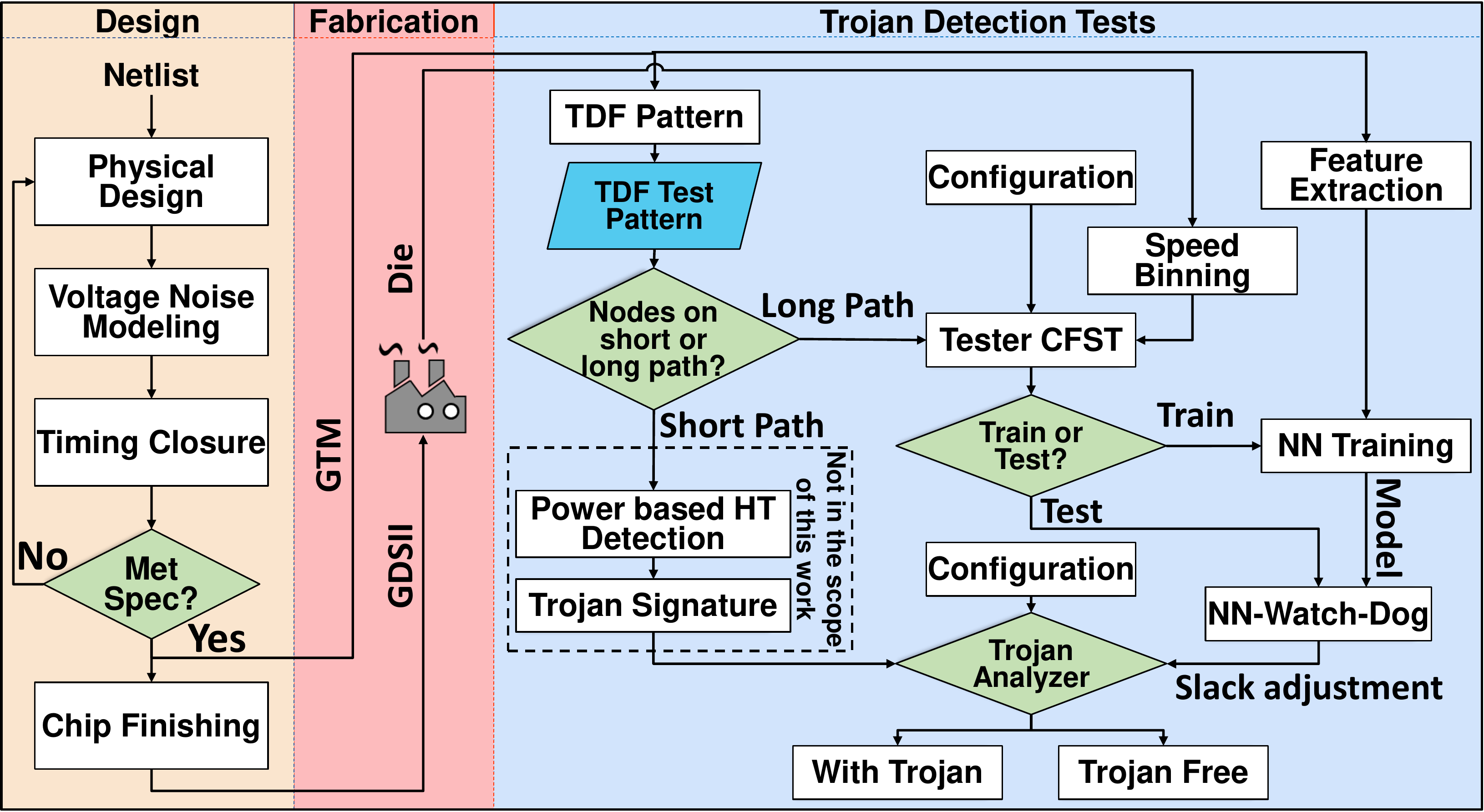}
\caption{LASCA Trojan Detection Flow: The model includes changes in the design and test stages. The test stage divides the timing-paths into long \& short paths. The short paths are subjected to power side-channel Trojan detection as described in \cite{lecomte2017chip} (not covered in this paper), and the long paths are subjected to delay side-channel analysis using GTM as reference timing model, adjusted by a NN that is trained as a process watchdog and by using CFST to find the start-to-fail frequencies for timing-paths under test.}
\label{flow_chart}
\end{figure}

Fig. \ref{flow_chart} shows the overall flow of the LASCA Trojan detection flow. We augment the design stage with an additional step for statistical modeling of the voltage noise and IR drop using IR-ATA flow as described in \cite{ASPDAC}. Accordingly, the STA reports the timing slack of each timing path based on its estimate of voltage drop and voltage noise (as opposed to a global pessimistic margin). This, as we will illustrate in the result section, will improve the correlation between timing slack predicted by timing engine, and the timing slack observed at test time using CFST. The final GDSII is then sent to the foundry for fabrication. The fabricated ICs may be tested in the untrusted foundry for functionality. The working ICs are then sent to a trusted facility for Trojan detection.

To detect a Trojan, we need to find the TT/TP induced slack change. As Fig. \ref{trojan_terminology} shows, a TT adds capacitive load to driving cell of its \textit{observed} net, and the TP appends an additional gate delay to every timing path that passes through its \textit{victim} net. To detect a victimized or a monitored net (by a TP or TT), and for having no prior knowledge on which nets are affected, we need to include all nets in our delay analysis. We define a P2P-wire as a net that connects the output pin of a driver cell (or a primary~input) to the input pin of one of its fanout cells (or a primary~output). Hence a gate with a fanout of 4 has 4 P2P-wires. Each P2P-wire will be tested for rise and fall transitions. To increase the detection rate and to account for PV, this process may be repeated for $N$ different timing-paths passing through that net. The second criteria for selecting the timing-paths is the maximum frequency of the tester equipment; The delay of the selected paths should be larger than the limit imposed by the maximum reachable frequency of the tester equipment. If the P2P-wire in no timing-path is long enough for CFST, it is regarded as a candidate for Trojan detection via power-based detection schemes. Note that timing-paths with a small number of gates (in their data sub-path) have high controllability, making them ideal for the power-based Trojan detection schemes (e.g. \cite{agrawal2007trojan, rad2010sensitivity,  salmani2009new, host_review}) that rely on full or partial activation of such paths. For all other timing-path candidates, we generate the Path Delay Fault (PDF) test vectors using an Automatic Test Pattern Generation tool (ATPG). If ATPG cannot generate a test pattern for a path, the path selection changes. If ATPG cannot generate a test vector for any path through that P2P-wire, it is discarded.  

\begin{algorithm}[!htb]
\caption{Generating a training set for the NN-Watchdog}\label{training_d}
\begin{algorithmic}[1]
\scriptsize
\State \textcolor{black}{$NP$} $\gets$ $mR^2$ \Comment{R is the registers count, and $m$ is a large number (e.g. 10)}
\State \textcolor{black}{$TimingPaths$} $\gets$ Select \textcolor{black}{$NP$} timing-paths (min of $m$ path per endpoint) 
\ForAll{\textcolor{black}{$path$} in \textcolor{black}{$TimingPaths$}}
    \State \textcolor{black}{feature($path$)} $\gets$ Extract \textcolor{black}{$path$} features from GTM \Comment{input feature}
    \State \textcolor{black}{GTM($path$)} $\gets$ Extract \textcolor{black}{$path$} slack from GTM  
    \State \textcolor{black}{$Slack($path$)$} = 0
\EndFor

\ForAll{\textcolor{black}{$die$} in \textcolor{black}{$Dies$}}
    \ForAll{\textcolor{black}{$path$} in \textcolor{black}{$TimingPaths$}}
        \State \textcolor{black}{CFST($die$,$path$)} $\gets$ Slack of \textcolor{black}{$path$} in CFST test of \textcolor{black}{$die$}
        \State \textcolor{black}{Slack($path$)} += \textcolor{black}{CFST($die$,$path$)}
    \EndFor
\EndFor
\ForAll{\textcolor{black}{$path$} in \textcolor{black}{$TimingPaths$}}
    \State \textcolor{black}{Slack($path$)} = \textcolor{black}{Slack($path$)/$NP$};
    \State \textcolor{black}{$\Delta_{slack}(path)$} = \textcolor{black}{Slack($path$)} - \textcolor{black}{GTM($path$)} \Comment{label}
    \State \textcolor{black}{data-points($path$)} $\gets$ \textcolor{black}{(features($path$),$\Delta_{slack}(path)$)}
\EndFor
\normalsize
\end{algorithmic}
\end{algorithm}

The Alg. \ref{Trojan_detection_flow} describes our proposed Trojan detection flow.  As described in this algorithm, after selecting the set of timing paths for PDF testing, we speed-bin the fabricated dies. In the next step, we collect the NN-Watchdog training data using the flow described in algorithm \ref{training_d}. Then, we train a process tracking NN-Watchdog for each bin and extract the standard deviation of each NN-Watchdog in predicting the shifted delays. For each bin, we perform CFST and measure the start to fail frequencies for the selected timing-paths. The slack difference  ($\delta$) between the mean of slacks reported by the CFST and the NN-Watchdog adjusted slack from GTM (in the same bin) represents the likelihood of a timing path being affected by a Trojan. To make a binary decision, we use a threshold to assess the significance of $\delta$ and classify the timing paths into benign or malignant (Trojan) classes. 

When choosing a value for Trojan-detection threshold, we face a trade-off between the false positive rate and the accuracy of Trojan detection. The false positive could be the result of 1) inaccuracy in the GTM, 2) inaccuracy of NN, and 3) random PV for sampling over a small number of ICs. To reduce the false positive rate, the threshold used for detection should be large enough, to account for these. Since we average the delay of each timing-path over many IC samples, the impact of random PV in the average delay could be reduced to a desirable range. However, we still have to account for the inaccuracy of the NN and persistent variation. Hence, we define the detection threshold to be $T_{Th} = n \times  max(\sigma_{NN}, \sigma_{PV})$, in which the $\sigma_{PV}$ is the expected variance of persistent PV (excluding random) and $\sigma_{NN}$ is the standard deviation of the NN. Since $\sigma_{NN}$ is the aggregated impact of NN inaccuracy (for under-fitting or over-fitting of the trained model) and impact of persistent PV, the variance of $\sigma_{NN}$ tends to be larger than $\sigma_{PV}$, and we can simply use $T_{Th} = n \times \sigma_{NN}$ (n is selected as 4 in Alg.~\ref{Trojan_detection_flow}).

To verify the choice of threshold values $T_{Th}$, we utilized Youden\cite{youden_index} method to extract the threshold value from a Receiver Operating Characteristic (ROC) curve that we generate over our SPICE simulation data (details in Section~\ref{results_section}). \textit{Note that at test time, we do not know which timing-paths are affected by HW Trojan. Hence, the optimal threshold of detection cannot be determined using the Youden method}. 

Change in the temperature affects the speed of transistors and alters the RC characteristics of the connecting wires. But, the temperature change is an extremely slow phenomenon. That's why one can design temperature sensors with sampling frequencies far lower than operational clock frequency \cite{5537410, 4512063}. At test time, a test vector is loaded into the scan chain using a slow clock, then the circuit operates at-speed for two cycles (launch and capture) using a fast clock. Finally, the scan is offloaded using a slow clock. The heat dissipation when using a slow clock is quite low, and the duration of at-speed test is only two cycles for each test pattern, limiting the extent of temperature changes to a fraction of a degree Celsius. Hence, at test time the die temperature can be tightly controlled to discount the delay impact of temperature~variations.

\begin{algorithm}[!htb]

\caption{LASCA Trojan Detection Flow }\label{Trojan_detection_flow}
\begin{algorithmic}[1]
\scriptsize

\State \textcolor{black}{$N$} = \# paths to be tested through each net in the design
\State \textcolor{black}{$Nets$} $\gets$ all nets in the design.
\ForAll{\textcolor{black}{$net$} in \textcolor{black}{$Nets$}} \Comment{net selection of Path Delay Fault (PDF) test}
    \State \textcolor{black}{$TimingPaths$} $+=$ select \textcolor{black}{$N$} timing-paths passing through \textcolor{black}{$net$} 
\EndFor

\State Perform speed binning on all dies and assign them to \textcolor{black}{B} bins. 

\ForAll{\textcolor{black}{$bin$} in \textcolor{black}{$B$}}  \Comment{NN training}
    \State \textcolor{black}{$NN_{bin}$} $\gets$ Train a NN-Watchdog according to the algorithm \ref{training_d}
    \State \textcolor{black}{$\sigma_{NN_{bin}}$} $\gets$ the standard deviation of \textcolor{black}{$NN_{bin}$}

    \ForAll{\textcolor{black}{$die$} in \textcolor{black}{$bin$}}
        \State \textcolor{black}{Slack} = 0
       
        \ForAll{\textcolor{black}{$path$} in \textcolor{black}{$TimingPaths$}}
            \State \textcolor{black}{CFST(bin,die,path)} $\gets$ \textcolor{black}{$path$} slack measured by CFST \textcolor{black}{$die$} in the \textcolor{black}{$bin$}
            \State \textcolor{black}{Slack(bin,path)} += \textcolor{black}{CFST(bin,die,path)}
        \EndFor   
    \EndFor
    
    \ForAll{\textcolor{black}{$path$} in \textcolor{black}{$TimingPaths$}}
        \State \textcolor{black}{$\mu_{S}$(bin,path)} = \textcolor{black}{Slack(bin,path)}/sizeof(\textcolor{black}{bin});
    \EndFor

\State \textcolor{black}{$T_{Th}$}= $4\times $\textcolor{black}{$\sigma_{NN_{bin}}$} \Comment{Detection Threshold $=4\sigma$ to reduce false positive}
\ForAll{\textcolor{black}{$path$} in \textcolor{black}{$TimingPaths$}}
    \State \textcolor{black}{GTM(path)} $\gets$ query the slack of \textcolor{black}{path} from GTM
    \State \textcolor{black}{NNSD(path)} $\gets$ slack shift suggested by \textcolor{black}{$NN_{bin}(path)$} 
    \State \textcolor{black}{AS(path)} = \textcolor{black}{GTM(path)} + \textcolor{black}{NN$_{Watchdog}$(path)} \Comment{Adjusted Slack}
    \State \textcolor{black}{$\delta$}  = \textcolor{black}{$\mu_{S}$(bin,path)} - \textcolor{black}{AS(path)} \Comment{Shifted delay after adjustment}
    \If{ (\textcolor{black}{$\delta$}  $>$ \textcolor{black}{$T_{Th}$)}}   \Comment{Trojan Classifier}
        \State Likely Trojan Set $\gets \textcolor{black}{path}$
    \EndIf
\EndFor
\EndFor
\normalsize
\end{algorithmic}
\end{algorithm}


\section{Results and Discussion}\label{results_section}

In this section, we first look at the accuracy of the NN-Watchdog in tracking the process drift, and then we present the result of applying our proposed test flow, LASCA, for Trojan detection.  

  \vspace{1mm}
\subsection{NN-Watchdog Accuracy}\label{nn_results_accuracy}

\begin{figure}[t]
\centering
\includegraphics[width=\columnwidth]{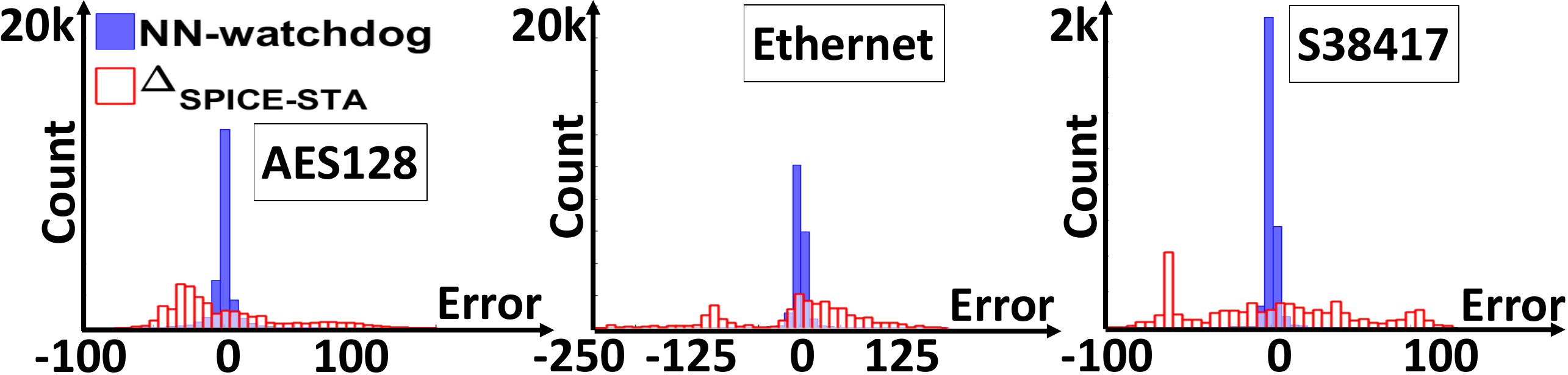}
\caption{Histogram of NN-Watchdog Error trained for different benchmarks.}
\label{NN_hist}
\end{figure}

Table \ref{NN_training_results} depicts the mean and standard deviation of the NN-Watchdog in predicting the shift in the delay of timing-paths when subjected to process drift. As shown, the standard deviation is reasonably small. To put this in perspective, we can compare the error distribution of NN-Watchdog with the error distribution obtained by finding the difference between delay of timing-paths reported by SPICE ($d_{SPICE}$) and that obtained from STA ($d_{STA}$). Fig. \ref{NN_hist} depicts the distribution of NN-Watchdog error and mean-shifted delay-difference model ($\Delta_{SPICE-STA} = d_{STA}-d_{SPICE})$ over a large selection of timing-paths.
As illustrated, the mean shifted SPICE-STA difference, for all benchmarks, has a much larger standard deviation compared to the NN-Watchdog error. This reveals the strength of NN-Watchdog and justifies why an NN-Watchdog could significantly enhance our Trojan detection flow by accurately adjusting the STA reported delay information to account for the impact of process drift.

\begin{table}[t]
\centering
\caption{The Accuracy of the NN-Watchdog regression model trained for different benchmarks. The $\mu$ and $\sigma$ are the Mean and Standard deviation of the regression error over the validation set. As discussed in Section \ref{NN_section}, the Fast, Typical and Slow process are simulated using skewed Spice model with (X,Y) = (5,5), (0,0), (-5,-5), respectively. }
\scalebox{0.75}{
\label{NN_training_results}
\begin{tabular}{|l|c|c|c|c|c|c|c|c|c|}
\hline
\multicolumn{1}{|c|}{\textbf{Benchmarks}} & 
\multicolumn{1}{|c|}{\textbf{Gates}} & 
\multicolumn{2}{c|}{\textbf{Size}}  & 
\multicolumn{2}{c|}{\textbf{Fast}}  &              
\multicolumn{2}{c|}{\textbf{Typical}}  &   
\multicolumn{2}{c|}{\textbf{Slow}}     
\\ \cline{3-10} 
\multicolumn{1}{|c|}{} & 
\multicolumn{1}{|c|}{} & 
\textbf{Train} & 
\textbf{Test} & 
\textbf{$\mu$(ps)}     & 
\textbf{$\sigma$(ps)}  &
\textbf{$\mu$(ps)}     & 
\textbf{$\sigma$(ps)}  &
\textbf{$\mu$(ps)}     & 
\textbf{$\sigma$(ps)} \\ \hline 
AES128             & 114K & 21K & 4K & -0.14 & 7.45 & 0.04  & 8.12  & -0.02 & 7.15  \\
Ethernet           & 40K  & 20K   & 4K   & 0.79   & 9.65 & 0.28  & 9.13  & -0.65  & 8.36  \\
S38417             & 6K   & 4K    & 1K   & 0.12   & 6.87 & 0.08  & 7.07  & 0.25   & 6.38  \\
\hline 
\end{tabular}
}
\end{table}

\subsection{HW Trojan Detection Accuracy}

\textbf{Setup:} We selected 720 timing-paths from non-critical to critical range,  covering a range of 400 ps of slack from 3 largest IWLS benchmarks \cite{iwls2} (Ethernet, S38417 and AES128). Each benchmark is hardened (physical design) and timing closed at 1.4GHz in 32nm technology. For each benchmark, we divided the selected timing-paths into two groups (360 each) for inserting TTs and TPs. We further divided each subgroup into three smaller groups of 120 paths each to implement small, medium, and large size Trojans. The TP size is controlled by the selection of logic gates with different inherent delays. The TT size is controlled by the distance it is placed from the triggering net. During NN-Watchdog training, we do not know if a timing-path selected for training contains a Trojan. Hence, we also evaluated the impact of including Trojans affected timing paths in the training; We trained 3 NN-Watchdogs with 0, 20 and 40 Trojan paths included in their training set. The rest of the Trojans are used for evaluating the LASCA Trojan detection accuracy as a part of its test-set. 

To model the voltage variation, we used Redhawk \cite{rh} and simulated 50 cycles of vectorless IR simulation when clock and data toggle rates are 100\% and 10\% respectively. In the SPICE simulation, each timing-path is assigned a random value from a normal distribution for the $V_{th}$ of its transistors (to model the PV), and each of its gates is annotated with the gate voltage reported by Redhawk in one simulation cycle. Note that each SPICE simulation presents a CFST test performed on a different die at a different time. Furthermore, the slack reported by the SPICE simulation for each timing-path was adjusted to the neighboring larger clock sweeping frequency step, modeling the CFST step size. The step size in the state-of-the-art tester equipments can be as small as 10-15ps. Hence, we selected the step size of the tester as 15ps.

In our simulations, we assessed the effectiveness of Trojan detection using 3 approaches. 1) Shifted STA (SSTA): when STA results are used as Golden Timing Model to detect HW Trojans. The process drift makes the direct usage of STA results quite ineffective. To account for process drift in SSTA, we have computed a static shift value, obtained from averaging the observed shift from many sampled timing-paths, and have shifted all reported slacks by STA using this value. For this approach, we have set the detection threshold to the fixed value of \textit{45ps} which is the delay of a 2-input NAND gate in our standard cell library. 2) Shifted GTM (SGTM): which is similar to SSTA with the exception that the voltage noise and IR-drop are modeled using IR-ATA \cite{ASPDAC} and the Trojan detection threshold is set to \textit{45ps}. 3) Neural shifted Golden Timing Model (NGTM) in which the voltage noise is modeled using IR-ATA \cite{ASPDAC}, while the process drift is modeled using NN-Watchdog. The NGTM represents the detection model proposed and utilized in LASCA. Furthermore, we have investigated the accuracy of NGTM when the training set includes 0, 20 and 40 timing-paths affected by HW Trojans. In all of SSTA, SGTM and NGTM, the effectiveness of the selected threshold is assessed by extracting the optimal threshold from the ROC curve using Youden\cite{youden_index} method. 

\begin{figure}[t]
\centering
\includegraphics[width=\columnwidth]{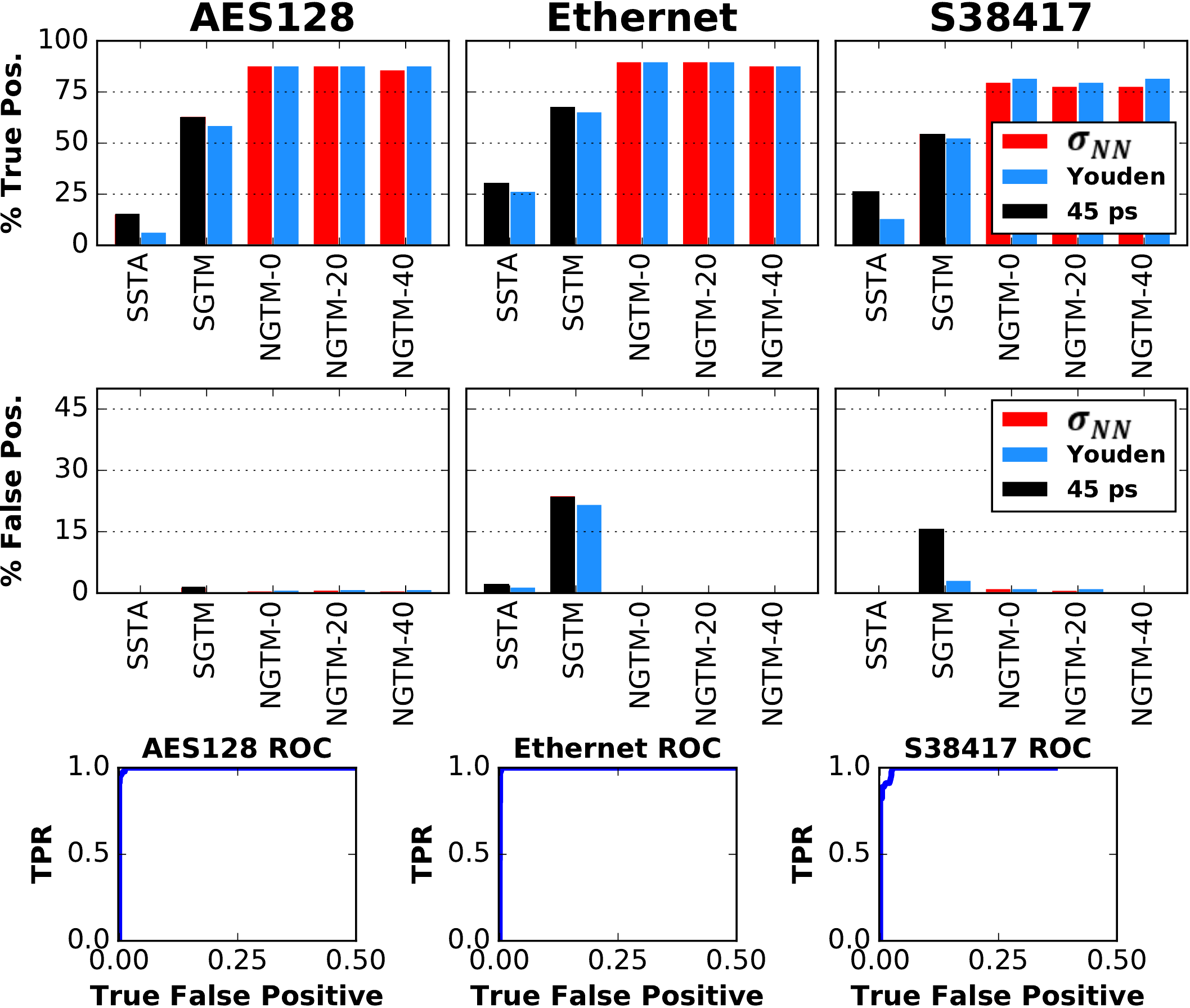}
\caption{Trojan Payload detection results for 3 benchmarks. (top): Detection rate, (middle): False positive rate, (bottom):
Associated ROC curve capturing the True Positive Rate (TPR) versus True False Positive Rate. The SSTA bar represents the HW Trojan Payload detection using a (Mean shifted) STA. The SGTM represents Trojan detection when IR-ATA\cite{ASPDAC} flow is deployed. The NGTM bars represent the Trojan Payload detection when both IR-ATA approach and the NN-Watchdog are combined. Each bar shows the NN trained when $X$ Trojans are included in the training set, with $X\in \{0,20,40\}$.}
\label{detect_TP_trojans}
\end{figure}

\begin{table}[t]
\centering
\caption{Threshold values used for TT and TP Trojan detection in Fast-bin in Algorithm \ref{Trojan_detection_flow}} \label{Threshold_optimal}
\scalebox{0.9}{
\begin{tabular}{|l|c|c|c|c|}
\hline
\multicolumn{1}{|c|}{\multirow{2}{*}{\textbf{Benchmarks}}} & \multicolumn{2}{c|}{\textbf{TP}}                  & \multicolumn{2}{c|}{\textbf{TT}}                \\ \cline{2-5} 
\multicolumn{1}{|c|}{}  & \textbf{Youden} & \textbf{$4 \times \sigma_{NN}$} & \textbf{Youden}     & \textbf{\textbf{$4 \times \sigma_{NN}$}} \\ \hline
\textbf{AES128}   & 27.1 & 29.86 & 16.3  & 29.86      \\ \hline
\textbf{Ethernet} & 35.5 & 38.67 & 15.4  & 38.67        \\ \hline
\textbf{S38417}   & 24.7  & 27.46  & 17.2  & 27.46        \\ \hline
\end{tabular}
}
\end{table}

Fig. \ref{detect_TP_trojans} captures the result of TP detection in Fast (X,Y)=(5,5) speed bin. The top row compares the accuracy of SSTA, SGTM, and NGTM in detecting TPs, and the middle row reports the false positive rate of detection for each model across different benchmarks. The NGTM model is reported 3 times, corresponding to a model having 0, 20 and 40 Trojan paths included in its training set. The bottom row illustrates the ROC curve from which the Youden threshold (as described in Section \ref{detection_flow_variation}) is extracted. The threshold values used for detection using each of these methods is reported in Table \ref{Threshold_optimal}. As illustrated, the usage of IR-ATA in SGTM model improves the TP detection rate compared to the SSTA at the expense of higher false positive. However, the use of IR-ATA and NN-Watchdog in the NGTM not only results in a significantly higher increase in the TP detection rate (to over 88\%), but also significantly depresses the false positive rate. This confirms the ability of NN-Watchdog in modeling the complicated, non-linear, path-specific shift of delays resulting from process drift. Finally, note that the presence of a small number of Trojans in the training set does not affect the accuracy of trained NN-Watchdog as the impact of a few samples in a large training set is statistically insignificant.

\begin{figure}[t]
\centering
\includegraphics[width=\columnwidth]{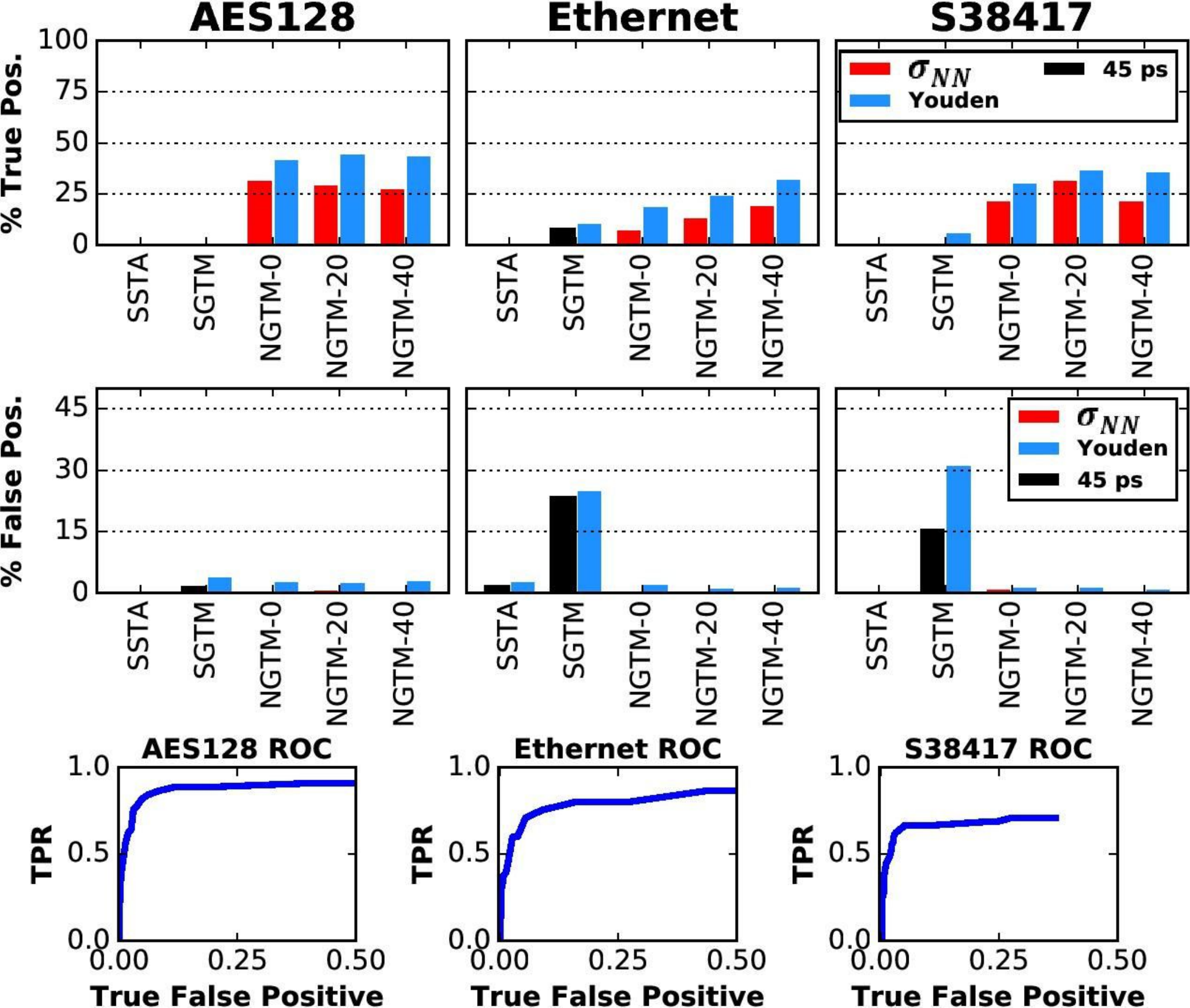}
\caption{Trojan Trigger detection results for 3 benchmarks. (top): Detection rate, (middle): False positive rate, (bottom): Associated ROC curve capturing the True Positive Rate (TPR) versus True False Positive Rate. \vspace{1mm}}
\label{detect_TT_trojans}
\end{figure}

\begin{table}[t]
\centering
\caption{Percentage of False Positives (FPo) and True Positives (TPo) when LASCA (as described in Alg. \ref{Trojan_detection_flow}) with NGTM-10 is used for detection of TP in Slow, Typical, and Fast speed bins.} \label{speed_bin_detection}
\scalebox{0.86}{
\begin{tabular}{|l|c|c|c|c|c|c|c|c|}
\hline
\multicolumn{1}{|c|}{\multirow{2}{*}{\textbf{Benchmarks}}} & 
\multicolumn{2}{c|}{\textbf{Slow}}      &
\multicolumn{2}{c|}{\textbf{Typical}}   &
\multicolumn{2}{c|}{\textbf{Fast}}   &
\multicolumn{2}{c|}{\textbf{No-Binning}}     \\ \cline{2-9} 
\multicolumn{1}{|c|}{}  & \textbf{TPo} & \textbf{FPo} & \textbf{TPo}     & \textbf{FPo} & \textbf{TPo}  & \textbf{FPo} & \textbf{TPo}  & \textbf{FPo}\\ \hline
\textbf{AES128}   & 88.6 & 0.11 & 87.8 & 0.17 & 86.1  &0.18  & 0.78 & 0.31  \\ \hline
\textbf{Ethernet} & 87.3 & 0.17 & 85.5 & 0.12 & 88.6  & 0.15    & 0.80 & 0.48  \\ \hline
\textbf{S38417}   & 83.7  & 0.19  & 82.2 & 0.23 & 80.3  & 0.39  & 0.77 & 0.45    \\ \hline
\end{tabular}
}
\end{table}

Figure~\ref{detect_TT_trojans} depicts the result of our TT detection in the FAST speed bin with (X,X) = (5,5). As shown, NGTM has a lower rate for detecting TTs compared to TPs due to the smaller impact of TT on the delay of affected observed nets compared to TP (which is at least equal to one gate delay). Similar to the TP case, we observe that contamination of the training set with few HW Trojan data points does not impact the accuracy of trained NN-Watchdog. This is because the number of HW Trojan infested timing paths is statistically insignificant and does not affect the training results. As illustrated, the Trojan trigger detection using our proposed approach closely tracks the Yuden model. Note that extracting the Yuden threshold requires a Trojan oracle database that is not available and is only presented to illustrate the effectiveness of our proposed solution. Finally, note that a hardware Trojan can have multiple TT and TP; although we have separately reported the result of TT and TP detection, detection of a single TT or TP is enough to detect the hardware Trojan. Therefore the overall detection rate of a hardware Trojan is larger than the results reported for TT or TP detection.

Table \ref{speed_bin_detection} captures the results of TP Trojan detection in all speed bins. As reported, the speed binning provides more accurate results for TP detection compared to the No-speed-binning case. This is due to the larger standard deviation of the NN-Watchdog when training over extracted delays from all dies without considering the impact of persistent PV.

\section{Conclusion}\label{conclusion_section}
In this paper, we presented LASCA, a promising methodology for Trojan detection based on side-channel delay analysis, that does not require the availability and usage of a Golden IC. For Trojan detection, The LASCA relies on 1) improving the timing model at design time to account for voltage noise, and 2) training a Neural Network at test time that is used as a process tracking watchdog to model the process drift (while accounting for process variation). We have reported that the LASCA Trojan detection flow could achieve close to $90\%$ Trojan detection in the selected benchmarks.

\nocite{synopsys}

\nocite{rh}

\scriptsize
\bibliographystyle{IEEEtran}
\bibliography{IEEEabrv,refs}

\end{document}